\documentclass[prb,a4paper,twocolumn,floatfix,showpacs,showkeys,amsmath,amssymb,nobibnotes,altaffilletter]{revtex4}

\usepackage{graphicx}
\usepackage{pslatex}
\usepackage{xspace}
\usepackage{subfigure}

\newcommand{\pht}{poly(3-hexyl thiophene)\xspace}
\newcommand{\pcbm}{[6,6]-phenyl-C$_{61}$ butyric acid methyl ester\xspace}

\begin{document}

\preprint{}

\title{Polaron Recombination in Pristine and Annealed Bulk Heterojunction Solar Cells}

\author{C.~Deibel}\email{deibel@physik.uni-wuerzburg.de}
\affiliation{Experimental Physics VI, Julius-Maximilians-University of W{\"u}rzburg, D-97074 W{\"u}rzburg}

\author{A.~Baumann}
\affiliation{Experimental Physics VI, Julius-Maximilians-University of W{\"u}rzburg, D-97074 W{\"u}rzburg}

\author{A.~Wagenpfahl}
\affiliation{Experimental Physics VI, Julius-Maximilians-University of W{\"u}rzburg, D-97074 W{\"u}rzburg}

\author{V.~Dyakonov}
\affiliation{Experimental Physics VI, Julius-Maximilians-University of W{\"u}rzburg, D-97074 W{\"u}rzburg}
\affiliation{Functional Materials for Energy Technology, Bavarian Centre for Applied Energy Research (ZAE Bayern), D-97074 W{\"u}rzburg}

\date{\today}

\begin{abstract}

The major loss mechanism of photogenerated polarons was investigated in P3HT:PCBM solar cells by the photo-CELIV technique. For pristine and annealed devices, we find that the experimental data can be explained by a bimolecular recombination rate reduced by a factor of about ten (pristine) and 25 (annealed) as compared to Langevin theory. Aided by a macroscopic device model, we discuss the implications of the lowered loss rate on the characteristics of polymer:fullerene solar cells.

\end{abstract}

\pacs{71.23.An, 72.20.Jv, 72.80.Le, 73.50.Pz, 73.63.Bd}

\keywords{organic semiconductors; polymers; photovoltaic effect; charge carrier recombination}

\maketitle

Organic bulk heterojunction (BHJ) solar cells have shown an increasing performance in the recent year, and also scientific progress concerning the fundamental understanding has been made~\cite{scherf2008book}. However, the dominant loss mechanism of the photocurrent is still under discussion. In polymer:fullerene solar cells, usually bimolecular recombination processes are observed by charge extraction techniques,\cite{mozer2005b} whereas the short circuit current of state-of-the-art devices shows a monomolecular signature.\cite{riedel2004} Koster et al. approached an explanation of this discrepancy by applying a macroscopic device model,\cite{koster2005b} stating that bimolecular recombination at short circuit accounted for only a few percent loss, therefore being latent in the current--voltage measurements. In order to contribute to this discussion, we apply photo-induced charge extraction by linearly increasing voltage (photo-CELIV)~\cite{juska2000} on pristine and annealed P3HT:PCBM (\pht:\pcbm) solar cells in order to investigate the polaron recombination dynamics. 

We prepared organic bulk heterojunction solar cells by spin coating 1:1 blends of poly[3-hexyl thiophene-2,5-diyl] (P3HT) with [6,6]-phenyl-C61 butyric acid methyl ester (PCBM),  20mg/ml dissolved in Chlorobenzene, on PEDOT:PSS covered ITO/glass substrates. The active layer was about 105nm thick. Al anodes were thermally evaporated. We obtained P3HT from Rieke Metals and PCBM from Solenne. The photo-CELIV method was applied on pristine and annealed samples. Charge carriers were generated by a short laser pulse (Nitrogen laser with dye unit, 5ns, 500$\mu$J/cm$^2$). After a delay time at zero internal field, the remaining charges are extracted by a voltage ramp in reverse bias. The charge carrier mobility and the concentration of extracted charge carriers are obtained simultaneously. Furthermore, we use a macroscopic simulation program implemented by us which solves the differential equation system of the Poisson, continuity and drift--diffusion equations by an iterative approach, as described in Ref.~\cite{deibel2008a}. 

\begin{figure}
	\includegraphics[width=7.5cm]{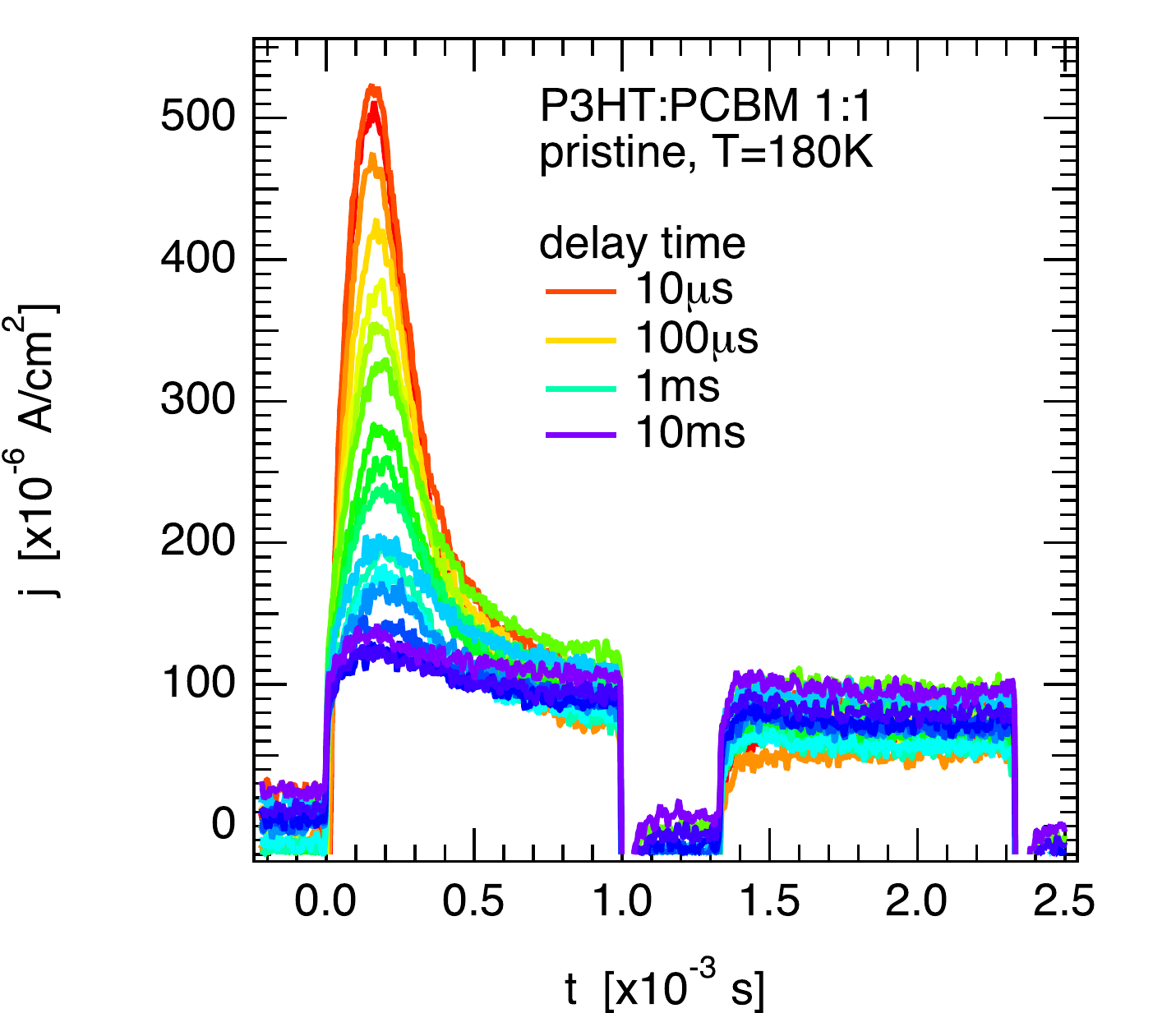}
	\caption{Photo-CELIV spectrum of a pristine P3HT:PCBM solar cell in dependence of the delay time betwen laser pulse and extraction voltage pulse at 180K. The second extraction peak shows the negligible influence of injection currents for the pristine sample.\label{fig:celiv_prist_180K}}
\end{figure}

A photo-CELIV measurement of a pristine P3HT:PCBM solar cell is shown in Fig.~\ref{fig:celiv_prist_180K}. The evaluation of the extracted charge concentration in dependence on the delay time gives us direct insight into the polaron recombination dynamics. Neither monomolecular nor Langevin-type bimolecular recombination are able to fit the experimental data well. Instead, the data is to be described by a reduced Langevin recombination rate 
\begin{equation}
	R = \zeta \gamma (n p - n_i^2 )
	\label{eqn:Langevin}
\end{equation}
with a prefactor $\zeta<1$ ($\zeta=1$ corresponds to the original Langevin theory~\cite{langevin1908}), where $n$ and $p$ are electron and hole concentration, respectively, $n_i$ is the intrinsic carrier concentration, and $\gamma$ is the Langevin recombination parameter. The latter is linearly proportional to the charge carrier mobility. We note that a trimolecular fit ($dn/dt \propto n^3$) gave an even lower deviation for the 180K pristine P3HT:PCBM sample. Despite a very recent report with similar findings~\cite{shuttle2008,juska2008private}, further investigations are necessary to verify this unexpected result. Therefore, and in analogy to literature~\cite{pivrikas2005,juska2006} and our previous results~\cite{deibel2008mrs} on annealed P3HT:PCBM devices only, we interpret the experimental data in view of reduced Langevin rates for both pristine and annealed samples. At 180K, the pristine solar cells show $\zeta$ of around 0.1, which is further reduced to $\zeta=0.04$ for annealed devices. At higher temperatures, the pristine sample stays at approximately 0.1. For the annealed sample, the determination of the prefactors is strongly influenced by charge injection, increasing the error margin; within its bounds, we see no temperature dependent variation of $\zeta$. Previously reported was an even weaker Langevin recombination rate, thus a lower $\zeta$, at higher temperatures~\cite{juska2006,deibel2008mrs}.

\begin{figure}
	\includegraphics[width=7.5cm]{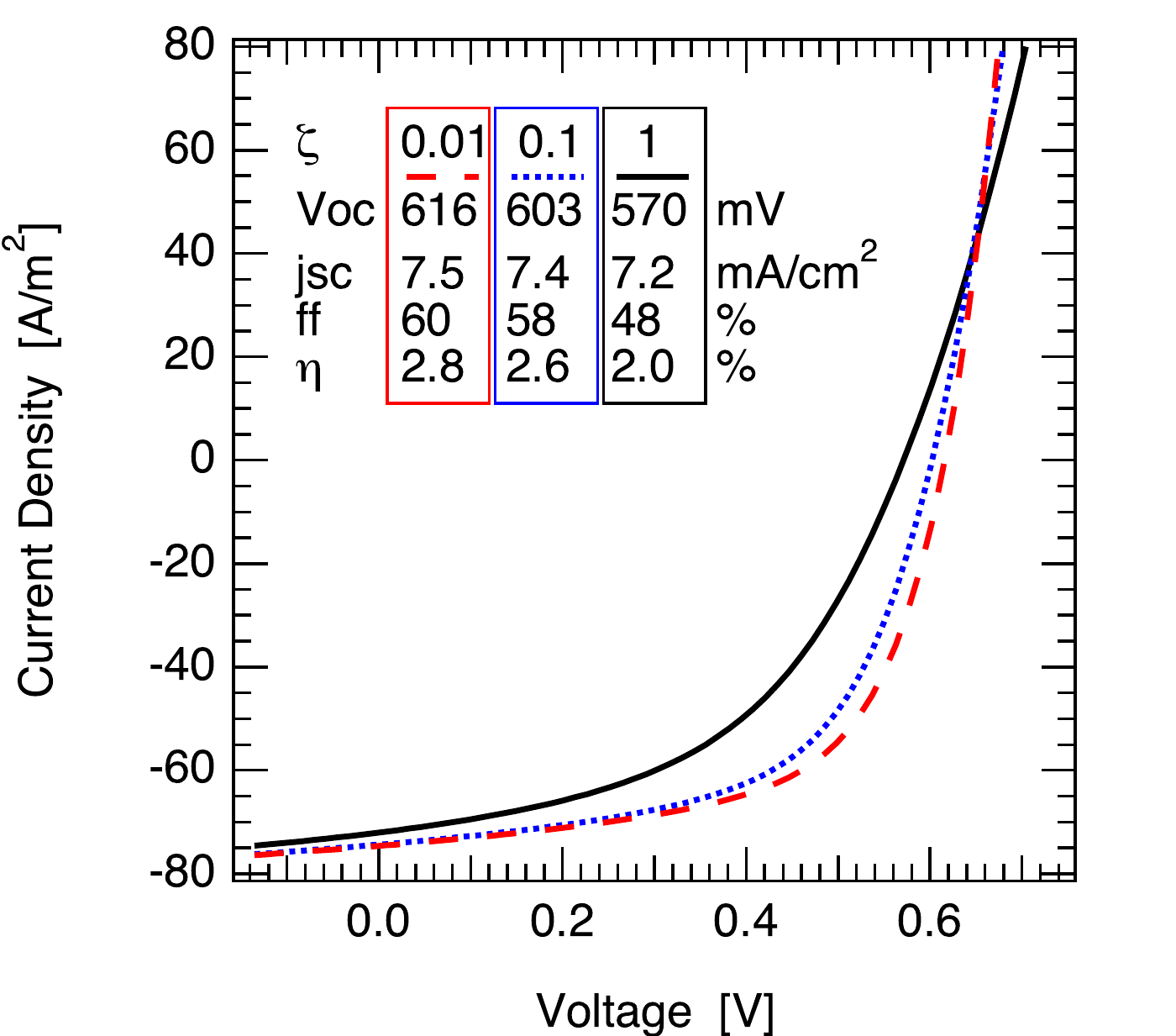}%
	\caption{Simulated current--voltage characteristics of a polymer:fullerene solar cell under one sun for three different values of the recombination prefactor $\zeta$. The pronounced influence of the reduced bimolecular recombination on the field-dependent photocurrent can be clearly seen.\label{fig:sim-iv}}
\end{figure}

\begin{figure}
	\includegraphics[width=7.5cm]{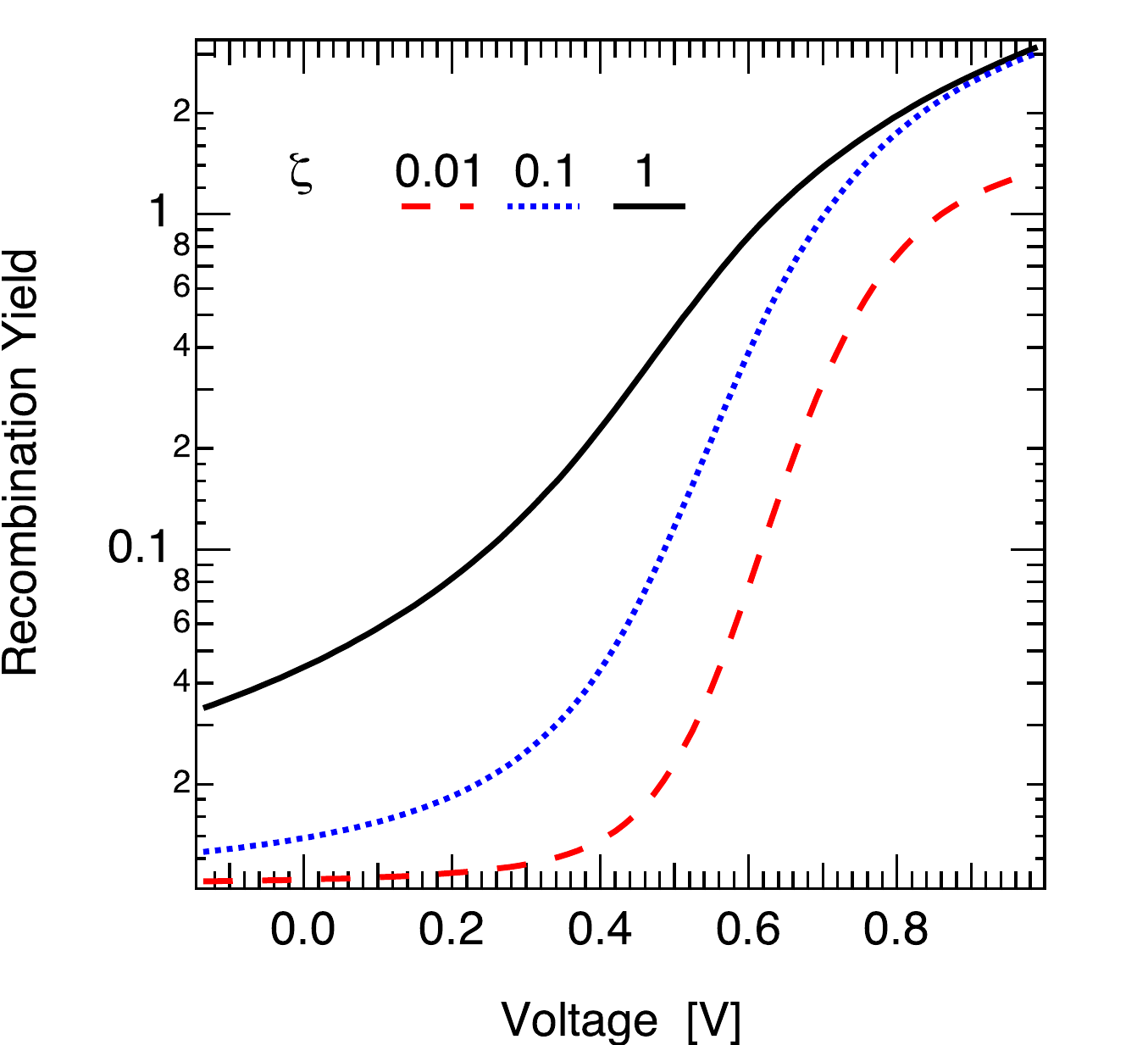}%
	\caption{Simulated bimolecular recombination yield (Eqn.~\ref{eqn:R}) of polymer:fullerene solar cells in dependence on the recombination pre\-factor $\zeta$.\label{fig:sim-rec}}
\end{figure}

In order to clarify the impact of a reduced bimolecular recombination rate on working polymer solar cells, we performed macroscopic simulations. The calculated current--voltage characteristics under one sun for three different values of $\zeta$, 1 (Langevin), 0.1, and 0.01, are shown in Fig.~\ref{fig:sim-iv} . Clearly, the field-dependent photocurrent is improved by a lowering of the bimolecular recombination rate. In order to quantify the influence of the latter, we use the recombination yield as a measure. It is defined as 
\begin{equation}
	\text{recombination yield} = 1 - \frac{U}{PG}
	\label{eqn:R}
\end{equation}
where $U=PG-(1-P)R$ is the net generation rate, $G$ is the exciton generation rate, and $P$ the polaron pair dissociation yield. The recombination yield is shown in Fig.~\ref{fig:sim-rec}. We note that the recombination term $R$ includes photogenerated and injected carriers, but is normalised to the polaron photogeneration rate $PG$. If normal Langevin recombination ($\zeta=1$) were applicable, even the short-circuit current would be lowered by bimolecular losses. With values of the prefactor of below a tenth as determined by our experiments, however, the losses start only at higher voltages. Thus, mainly the fill factor and the open-circuit voltage are negatively affected by the nongeminate recombination. The low recombination rate, however, limits its impact on the solar cell performance, so that the charge extraction properties are governing the solar cell efficiency rather than the bimolecular recombination.

\begin{figure}
	\includegraphics[width=7.5cm]{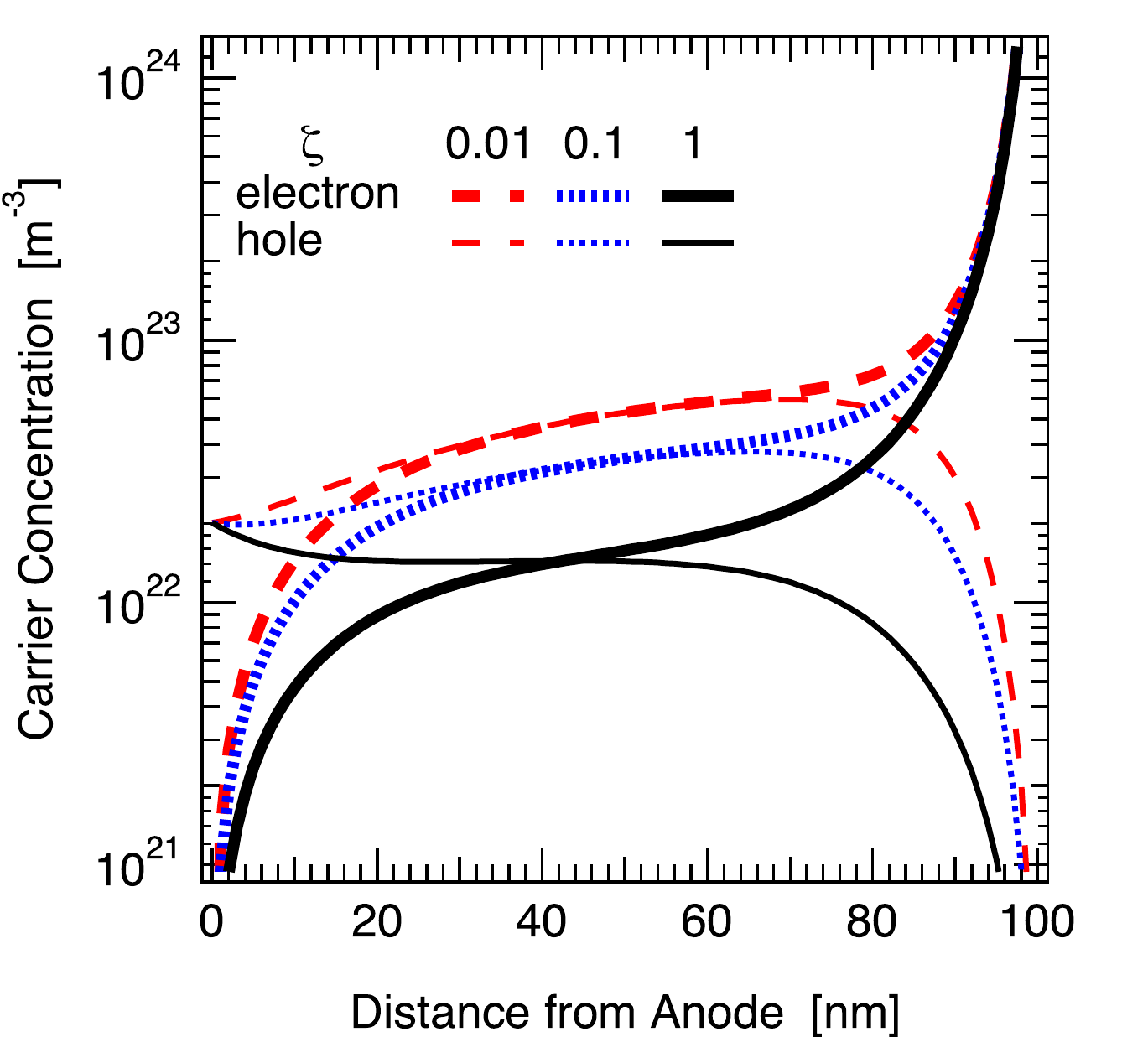}%
	\caption{Simulated carrier concentration in the active layer of polymer:fullerene solar cells under open-circuit conditions. The reduced bimolecular recombination leads to a significant increase of the carrier concentration.\label{fig:sim-n}}
\end{figure}

The effect of the recombination yield on the carrier concentration under open-circuit conditions is depicted in Fig.~\ref{fig:sim-n}. The lower the recombination rate, the higher the steady-state carrier concentration in the active area of the polymer:fullerene solar cell. Actually, this difference in carrier concentration is directly reflected in the solar cell characteristics: the higher carrier concentration indicates that the quasi-Fermi levels move closer to their respective bands. This leaves more room for the open-circuit voltage, which is thus increased for raised electron and hole concentrations due to lower nongeminate loss rates.

In conclusion, by performing charge extraction experiments on pristine and annealed P3HT:PCBM solar cells, we have shown that the bimolecular loss rate is reduced significantly as compared to Langevin theory. The fill factor and the open-circuit voltage are reduced by the nongeminate recombination, but due to the low recombination rate, the solar cell performance is mainly determined by the charge extraction properties.
Nevertheless, the finding of a reduced Langevin recombination rate has an impact on the understanding and modelling of polymer solar cells, as was shown by performing macroscopic device simulations, and therefore needs to be accounted for in future studies.

\begin{acknowledgments}

V.D.'s work at the ZAE Bayern is financed by the Bavarian Ministry of Economic Affairs, Infrastructure, Transport and Technology.

\end{acknowledgments}



\begin{thebibliography}{12}
\expandafter\ifx\csname natexlab\endcsname\relax\def\natexlab#1{#1}\fi
\expandafter\ifx\csname bibnamefont\endcsname\relax
  \def\bibnamefont#1{#1}\fi
\expandafter\ifx\csname bibfnamefont\endcsname\relax
  \def\bibfnamefont#1{#1}\fi
\expandafter\ifx\csname citenamefont\endcsname\relax
  \def\citenamefont#1{#1}\fi
\expandafter\ifx\csname url\endcsname\relax
  \def\url#1{\texttt{#1}}\fi
\expandafter\ifx\csname urlprefix\endcsname\relax\def\urlprefix{URL }\fi
\providecommand{\bibinfo}[2]{#2}
\providecommand{\eprint}[2][]{\url{#2}}

\bibitem[{\citenamefont{Scherf et~al.}(2008)\citenamefont{Scherf, Brabec, and
  Dyakonov}}]{scherf2008book}
\bibinfo{editor}{\bibfnamefont{U.}~\bibnamefont{Scherf}},
  \bibinfo{editor}{\bibfnamefont{C.}~\bibnamefont{Brabec}}, \bibnamefont{and}
  \bibinfo{editor}{\bibfnamefont{V.}~\bibnamefont{Dyakonov}}, eds.,
  \emph{\bibinfo{title}{Organic Photovoltaics. Materials, Device Physics, and
  Manufacturing Technologies}} (\bibinfo{publisher}{Wiley VCH},
  \bibinfo{year}{2008}).

\bibitem[{\citenamefont{Mozer et~al.}(2005)\citenamefont{Mozer, Dennler,
  Sariciftci, Westerling, Pivrikas, {\"O}sterbacka, and
  Ju{\v{s}}ka}}]{mozer2005b}
\bibinfo{author}{\bibfnamefont{A.~J.} \bibnamefont{Mozer}},
  \bibinfo{author}{\bibfnamefont{G.}~\bibnamefont{Dennler}},
  \bibinfo{author}{\bibfnamefont{N.~S.} \bibnamefont{Sariciftci}},
  \bibinfo{author}{\bibfnamefont{M.}~\bibnamefont{Westerling}},
  \bibinfo{author}{\bibfnamefont{A.}~\bibnamefont{Pivrikas}},
  \bibinfo{author}{\bibfnamefont{R.}~\bibnamefont{{\"O}sterbacka}},
  \bibnamefont{and}
  \bibinfo{author}{\bibfnamefont{G.}~\bibnamefont{Ju{\v{s}}ka}},
  \bibinfo{journal}{Phys. Rev. B} \textbf{\bibinfo{volume}{72}},
  \bibinfo{pages}{035217} (\bibinfo{year}{2005}).

\bibitem[{\citenamefont{Riedel et~al.}(2004)\citenamefont{Riedel, Parisi,
  Dyakonov, Lutsen, Vanderzande, and Hummelen}}]{riedel2004}
\bibinfo{author}{\bibfnamefont{I.}~\bibnamefont{Riedel}},
  \bibinfo{author}{\bibfnamefont{J.}~\bibnamefont{Parisi}},
  \bibinfo{author}{\bibfnamefont{V.}~\bibnamefont{Dyakonov}},
  \bibinfo{author}{\bibfnamefont{L.}~\bibnamefont{Lutsen}},
  \bibinfo{author}{\bibfnamefont{D.}~\bibnamefont{Vanderzande}},
  \bibnamefont{and} \bibinfo{author}{\bibfnamefont{J.~C.}
  \bibnamefont{Hummelen}}, \bibinfo{journal}{Adv. Funct. Mater.}
  \textbf{\bibinfo{volume}{14}}, \bibinfo{pages}{38} (\bibinfo{year}{2004}).

\bibitem[{\citenamefont{Koster et~al.}(2005)\citenamefont{Koster, Smits,
  Mihailetchi, and Blom}}]{koster2005b}
\bibinfo{author}{\bibfnamefont{L.~J.~A.} \bibnamefont{Koster}},
  \bibinfo{author}{\bibfnamefont{E.~C.~P.} \bibnamefont{Smits}},
  \bibinfo{author}{\bibfnamefont{V.~D.} \bibnamefont{Mihailetchi}},
  \bibnamefont{and} \bibinfo{author}{\bibfnamefont{P.~W.~M.}
  \bibnamefont{Blom}}, \bibinfo{journal}{Phys. Rev. B}
  \textbf{\bibinfo{volume}{72}}, \bibinfo{pages}{085205}
  (\bibinfo{year}{2005}).

\bibitem[{\citenamefont{Ju{\v{s}}ka et~al.}(2000)\citenamefont{Ju{\v{s}}ka,
  Arlauskas, Vili{\={u}}nas, and Ko{\v{c}}ka}}]{juska2000}
\bibinfo{author}{\bibfnamefont{G.}~\bibnamefont{Ju{\v{s}}ka}},
  \bibinfo{author}{\bibfnamefont{K.}~\bibnamefont{Arlauskas}},
  \bibinfo{author}{\bibfnamefont{M.}~\bibnamefont{Vili{\={u}}nas}},
  \bibnamefont{and}
  \bibinfo{author}{\bibfnamefont{J.}~\bibnamefont{Ko{\v{c}}ka}},
  \bibinfo{journal}{Phys. Rev. Lett.} \textbf{\bibinfo{volume}{84}},
  \bibinfo{pages}{4946} (\bibinfo{year}{2000}).

\bibitem[{\citenamefont{Deibel et~al.}(2008{\natexlab{a}})\citenamefont{Deibel,
  Wagenpfahl, and Dyakonov}}]{deibel2008a}
\bibinfo{author}{\bibfnamefont{C.}~\bibnamefont{Deibel}},
  \bibinfo{author}{\bibfnamefont{A.}~\bibnamefont{Wagenpfahl}},
  \bibnamefont{and} \bibinfo{author}{\bibfnamefont{V.}~\bibnamefont{Dyakonov}},
  \bibinfo{journal}{phys. stat. sol. (RRL)} \textbf{\bibinfo{volume}{1--3}}
  (\bibinfo{year}{2008}{\natexlab{a}}).

\bibitem[{\citenamefont{Langevin}(1909)}]{langevin1908}
\bibinfo{author}{\bibfnamefont{P.}~\bibnamefont{Langevin}},
  \bibinfo{journal}{C. R. Acad. Sci.} \textbf{\bibinfo{volume}{146}},
  \bibinfo{pages}{530} (\bibinfo{year}{1909}), \bibinfo{note}{translated by
  D.S.~Lemons and A.~Gythiel, Am. J. Phys. \textbf{65}, 1079 (1997).}

\bibitem[{\citenamefont{Ju{\v{s}}ka}(2008)}]{juska2008private}
\bibinfo{author}{\bibfnamefont{G.}~\bibnamefont{Ju{\v{s}}ka}},
  \bibinfo{journal}{private communications}  (\bibinfo{year}{2008}).

\bibitem[{\citenamefont{Shuttle et~al.}(2008)\citenamefont{Shuttle, O'Regan,
  Ballantyne, Nelson, Bradley, Mello, and Durrant}}]{shuttle2008}
\bibinfo{author}{\bibfnamefont{C.}~\bibnamefont{Shuttle}},
  \bibinfo{author}{\bibfnamefont{B.}~\bibnamefont{O'Regan}},
  \bibinfo{author}{\bibfnamefont{A.}~\bibnamefont{Ballantyne}},
  \bibinfo{author}{\bibfnamefont{J.}~\bibnamefont{Nelson}},
  \bibinfo{author}{\bibfnamefont{D.}~\bibnamefont{Bradley}},
  \bibinfo{author}{\bibfnamefont{J.~D.} \bibnamefont{Mello}}, \bibnamefont{and}
  \bibinfo{author}{\bibfnamefont{J.}~\bibnamefont{Durrant}},
  \bibinfo{journal}{Appl. Phys. Lett.} \textbf{\bibinfo{volume}{92}},
  \bibinfo{pages}{093311} (\bibinfo{year}{2008}).

\bibitem[{\citenamefont{Ju{\v{s}}ka et~al.}(2006)\citenamefont{Ju{\v{s}}ka,
  Arlauskas, Stuchlik, and {\"O}sterbacka}}]{juska2006}
\bibinfo{author}{\bibfnamefont{G.}~\bibnamefont{Ju{\v{s}}ka}},
  \bibinfo{author}{\bibfnamefont{K.}~\bibnamefont{Arlauskas}},
  \bibinfo{author}{\bibfnamefont{J.}~\bibnamefont{Stuchlik}}, \bibnamefont{and}
  \bibinfo{author}{\bibfnamefont{R.}~\bibnamefont{{\"O}sterbacka}},
  \bibinfo{journal}{J. Non-Cryst. Sol.} \textbf{\bibinfo{volume}{352}},
  \bibinfo{pages}{1167} (\bibinfo{year}{2006}).

\bibitem[{\citenamefont{Pivrikas et~al.}(2005)\citenamefont{Pivrikas,
  {\"O}sterbacka, Ju{\v{s}}ka, Arlauskas, and Stubb}}]{pivrikas2005}
\bibinfo{author}{\bibfnamefont{A.}~\bibnamefont{Pivrikas}},
  \bibinfo{author}{\bibfnamefont{R.}~\bibnamefont{{\"O}sterbacka}},
  \bibinfo{author}{\bibfnamefont{G.}~\bibnamefont{Ju{\v{s}}ka}},
  \bibinfo{author}{\bibfnamefont{K.}~\bibnamefont{Arlauskas}},
  \bibnamefont{and} \bibinfo{author}{\bibfnamefont{H.}~\bibnamefont{Stubb}},
  \bibinfo{journal}{Synth. Met.} \textbf{\bibinfo{volume}{155}},
  \bibinfo{pages}{242} (\bibinfo{year}{2005}).

\bibitem[{\citenamefont{Deibel et~al.}(2008{\natexlab{b}})\citenamefont{Deibel,
  Baumann, Lorrmann, and Dyakonov}}]{deibel2008mrs}
\bibinfo{author}{\bibfnamefont{C.}~\bibnamefont{Deibel}},
  \bibinfo{author}{\bibfnamefont{A.}~\bibnamefont{Baumann}},
  \bibinfo{author}{\bibfnamefont{J.}~\bibnamefont{Lorrmann}}, \bibnamefont{and}
  \bibinfo{author}{\bibfnamefont{V.}~\bibnamefont{Dyakonov}}, in
  \emph{\bibinfo{booktitle}{Mat. Res. Soc. Symp. Proc.}}
  (\bibinfo{year}{2008}{\natexlab{b}}).

\end{thebibliography}

\end{document}